%% file: Artigo.tex
\documentclass[3p,twocolumn,lefttitle]{elsarticle}
\usepackage{cmap}
\usepackage[utf8]{inputenc}
\usepackage{caption}
\usepackage{lineno,hyperref}
\usepackage{multicol}
\usepackage{multirow}
\usepackage{lscape}
\usepackage[T1]{fontenc}	
\usepackage[table,xcdraw]{xcolor}
\usepackage{float}
\usepackage[export]{adjustbox}
\usepackage{graphicx}
\usepackage{caption}
\usepackage{subcaption}
\usepackage{blindtext}
\usepackage{stfloats}
\usepackage{lipsum}
\usepackage{listliketab}
\usepackage{epstopdf}
\usepackage{numcompress}
\modulolinenumbers[5]

\journal{Journal of \LaTeX\ Templates}

\biboptions{sort&compress}
\begin{document}

\begin{frontmatter}

\title{
Evaluating the Efficacy of Vectocardiographic and ECG Parameters for Efficient Tertiary Cardiology Care Allocation Using Decision Tree Analysis}

\fntext[fn1]{Abbreviations: GEH: Global Electric Heterogeneity; ECG: Electrocardiogram; MI: Myiocardial Infarction; PCI: Percutaneous Intervention; CS: Cardiac Surgery; AUCPR: area under the Precision-Recall curve; AUC: Area Under the Curve;  SVG: Spatial Ventricular Gradient;VCG: Vectocardiogram; IQR: Interquartile interval }

\author[1]{Lucas José da Costa}

\author[2]{Vinicius Ruiz Uemoto}

\author[2]{Mariana F. N. de Marchi}

\author[2]{Renato de Aguiar Hortegal}

 \author[4]{Renata Valeri de Freitas\corref{mycorrespondingauthor}}
 \cortext[mycorrespondingauthor]{Corresponding author}
 \ead{renata.valfreitas@usp.br}

\affiliation[1]{organization={Instituto Internacional de Neuroci\^{e}ncia  
 Edmond  e  Lily  Safra},
city={Maca\'{i}ba},
country={Brazil}}

\affiliation[2]{organization={Instituto Dante Pazzanese de Cardiologia},
city={S\~{a}o Paulo},
country={Brazil}}

\affiliation[4]{organization={Escola Polit\'{e}cnica da Universidade de S\~{a}o Paulo},
city={S\~{a}o Paulo},
country={Brazil}}

\begin{abstract}

\textit{Objective:} Use real word data to evaluate the performance of the  electrocardiographic markers of GEH\footnote{Abbreviations: ECG: Electrocardiogram; VCG: Vectocardiogram; AUC: Area Under the Curve; AUCPR: area under the Precision-Recall curve; GEH: Global Electric Heterogeneity; SVG: Spatial Ventricular Gradient; MI: Myiocardial Infarction;PCI: Percutaneous Intervention; CS: Cardiac Surgery} as features in a machine learning model with Standard ECG features and Risk Factors in Predicting Outcome of patients in a population referred to a tertiary cardiology hospital. 

\textit{Methods:}  
Patients forwarded to specific evaluation in a cardiology specialized hospital performed an ECG and a risk factor anamnesis. A series of follow up attendances occurred in  periods of 6 months, 12 months and 15 months to check for cardiovascular related events (mortality or new nonfatal cardiovascular events (Stroke, MI, PCI, CS), as identified during 1-year phone follow-ups. 

The first attendance ECG was measured by a specialist and processed in order to obtain the global electric heterogeneity (GEH) using the Kors Matriz. The ECG measurements, GEH parameters and risk factors were combined for training multiple instances of XGBoost decision trees models. Each instance were optmized for the AUCPR and the instance with higher AUC is chosen as representative to the model. The importance of each parameter for the winner tree model was compared to better understand the improvement from using GEH parameters.

\textit{Results:}  The GEH parameters turned out to have statistical significance for this population specially the QRST angle and the SVG. The combined model with the tree parameters class had the best performance.
\textit{Conclusion:} The findings suggest that using VCG features can facilitate more accurate identification of patients who require tertiary care, thereby optimizing resource allocation and improving patient outcomes. Moreover, the decision tree model's transparency and ability to pinpoint critical features make it a valuable tool for clinical decision-making and align well with existing clinical practices.

\end{abstract}


      \end{frontmatter}

\section{Introduction}


Cardiovascular diseases are the main cause of death around the world \cite{who}. In 2022, 28\% of the deaths in Brazil had cardiovascular causes \cite{Relatorio2022}. Being able to detect and predict cardiovascular events, since one third of the deaths due to heart attacks and strokes occur in people under 70 years \cite{OMS-cardio}, is extremely positive and life-saving, particularly when it is done using non-invasive exams. 

The past decade was full of technological advances, specially the advances on computation and machine learning, with some of these having similar or better accuracy than specialists when diagnosing a disease. However a lot still need to be done in order to improve the quality of the models and their validation in clinical practice \cite{topol2019high}. 

The importance of ECG features when predicting cardiovascular outcomes was pointed by previous studies \cite{bouzid2021novel, MINCHOLE2019S61} as well as using ECG markers of global electric heterogeneity added to clinical characteristics can improve the detection of heart disease.  
\cite{waks2016global}\cite{Tereshchenko2018}\cite{Vondrak2022}

Therefore, it would  be interesting to evaluate the contribution that the GEH markers could give in a different scenario. This analysis was done using standard statistical tools and a machine learning model in order to evaluate the relevance of the GEH models in predicting the patient’s outcome. The goal is to provide an improved diagnostic tool having a high Sensitivity with the biggest sensibility possible. Therefore, no patient with a severe condition will be left with no medical care and the use of the resources of the health system will be optimized.

\section{Materials and Methods}
\subsection*{Population}

This study was performed in a population attended from May to August 2017 in a first attendance ambulatory in Dante Pazzanese Institute of Cardiology (IDPC), a tertiary/quaternary public healthcare center in São Paulo. Local institutional and national review boards approved the data collection protocols(CAAE 76085317.5.000.5462). The ambulatory in question is focused in triage patients referred from primary and secondary care centers. 

All patients that attended the ambulatory during the specified time period were contacted through phone calls during the follow ups and all patients that didn't answer it for any of the follow-ups (and therefore the outcome was unavailable) were excluded from the study's population. 303 subjects underwent a first attendance; after the 6 months follow-up 293 subjects' outcomes were confirmed by phone-call.  For the 1 year follow-up 274 subjects' outcomes were confirmed by phone-call.

The study population comprised 274 individuals, with a median age of 59 years (interquartile interval: 51.0 to 67.0) and a body mass index of 26.9 (IQR: 24.2 to 30.4). Of these individuals, 52.2\% were female. In terms of comorbidities, 69.0\% of the individuals presented with hypertension, emphasizing the high prevalence of this condition within the study population. Moreover, 27.7\% of the participants had diabetes.
Within this cohort, it is noteworthy that a significant proportion had experienced previous cardiac events. Specifically, 28.1\% of the participants had a documented history of Previous Myocardial Infarction (MI), indicating a considerable burden of ischemic heart disease. Additionally, 14.6\% had undergone previous Percutaneous Coronary Intervention (PCI), highlighting the need for coronary revascularization procedures. Furthermore, 11.3\% had a history of Previous Cardiac Surgery, reflecting the prevalence of complex cardiovascular conditions needing surgical interventions.

These findings emphasize the importance of targeted triage and comprehensive management strategies for individuals with cardiovascular diseases, particularly considering the substantial prevalence of risk factors such as hypertension and diabetes, as well as the notable occurrence of prior cardiac events in this population.

\subsection*{Data Collection}

All data from each patient was collected from 2 main sources: an 12-lead ECG exam and the patients previous cardiovascular events' history. The electrocardiography signals were obtained in 7 seconds acquisition of a 12-lead exam at a sampling rate of 240Hz, the traces were stored so they could be properly pre-processed and every exam was analyzed by a cardiology specialist that was also responsible for measuring the standard parameters 
from the ECG, as well as marking the baseline and peaks necessary for GEH computation. Patients' background was obtained during the subject's anamnesis and all data was saved in table format.
\begin{figure}[h!]
    \centering
    \includegraphics[trim=1.2cm 0cm 0cm 0cm, width=1\textwidth,height=17cm,keepaspectratio]{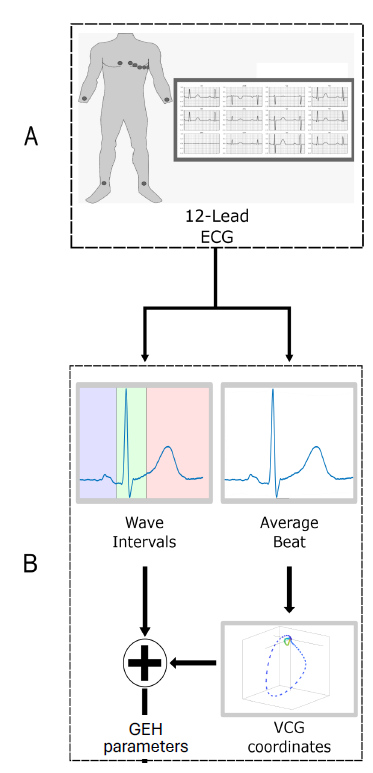}
    \caption{Process for obtaining the GEH parameters. 12-lead ECG (A), measurement of the wave intervals and average beat computation. Kors matrix is applied to generate the Frank leads and Vectocardiogram (B) Computation of the vectocardiographic features and Spatial Vector Gradient (C)}
    \label{Resumo}
\end{figure}

\begin{figure}[h]
    \centering
    \includegraphics[width=.4\textwidth,keepaspectratio]{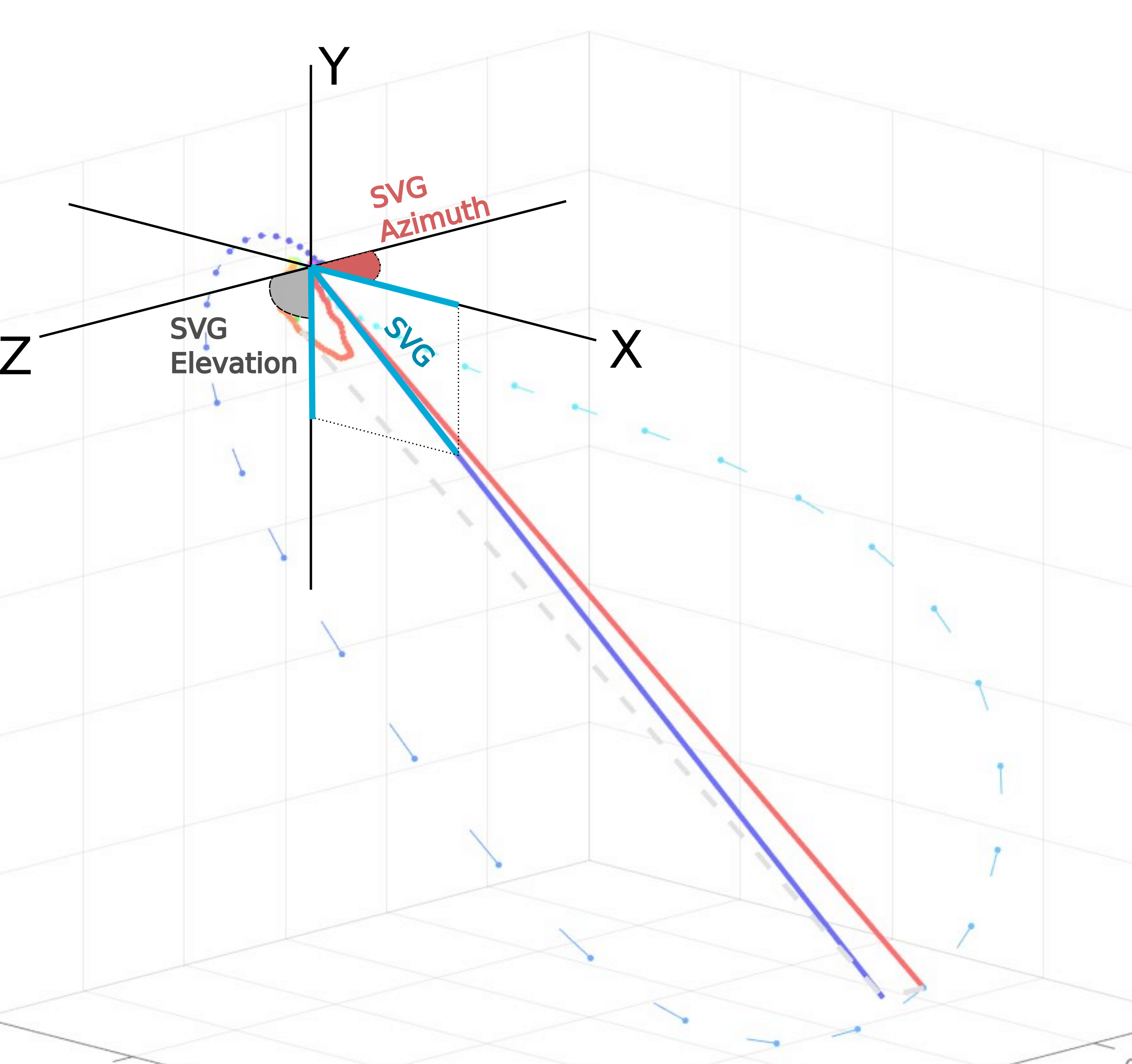}
    \caption{Representation of SVG Elevation and Azimuth angles.}
    \label{fig:SVG}
\end{figure} 
\subsection*{Data Pre-processing}

The raw ECG traces needed to be pre-processed so that they could be used to estimate the patient's vectorcardiogram through the Kors' transformation method \cite{jaros2019comparison}. The VCG would then be used to calculate the Global Electric Heterogeneity markers as was done by Waks \textit{et al}\cite{waks2016global}.

The ECG waves intervals were measured by a cardiology specialist using an in house developed ECG specialized software that marked the timestamp of the baseline, beginning, peak and ending of each wave of, at least, 3 cardiac cycles. Then, the median timestamp were submitted to an adapted version Tereshchenko's GEH Analysis algorithm\cite{GithubTereshchenko} that converts the ECG to VCG and then extract the GEH parameters (Figures \ref{Resumo} and \ref{fig:SVG}).

\subsection*{Parameter sets}
To investigate the effect of each type of parameter, we trained 4 machine learning models with different parameters sets. Three of those models contained only one category of features: Risk factors (R), Standard Electrocardiographic measurements (S) or Global heterogeneity measurements (G). The first three models serve as performance reference to how well a model with only that parameters set could perform while the fourth model (SRG) is a combinations of all the categories to evaluate if knowing the GEH parameters could lead to better predictions.

\subsection*{XGBoost}
The choice to use a decision tree with XGBoost as machine learning model is due to its simple architecture and ease of interpretation of results while having a great performance especially when working with tabular data\cite{XGB_tabular,shwartz2022tabular}.

Data were randomly divided into 70\% for training and 30\% for test. Due to the training data being unbalanced by a ratio of 5 negative outcomes to 1 positive, they needed to pass by both bootstrap oversampling and under-sampling before the learning process in order to avoid biases.

\begin{figure*}[h!]
    \centering
    \includegraphics[width=.95\textwidth,height=.9\textheight,keepaspectratio]{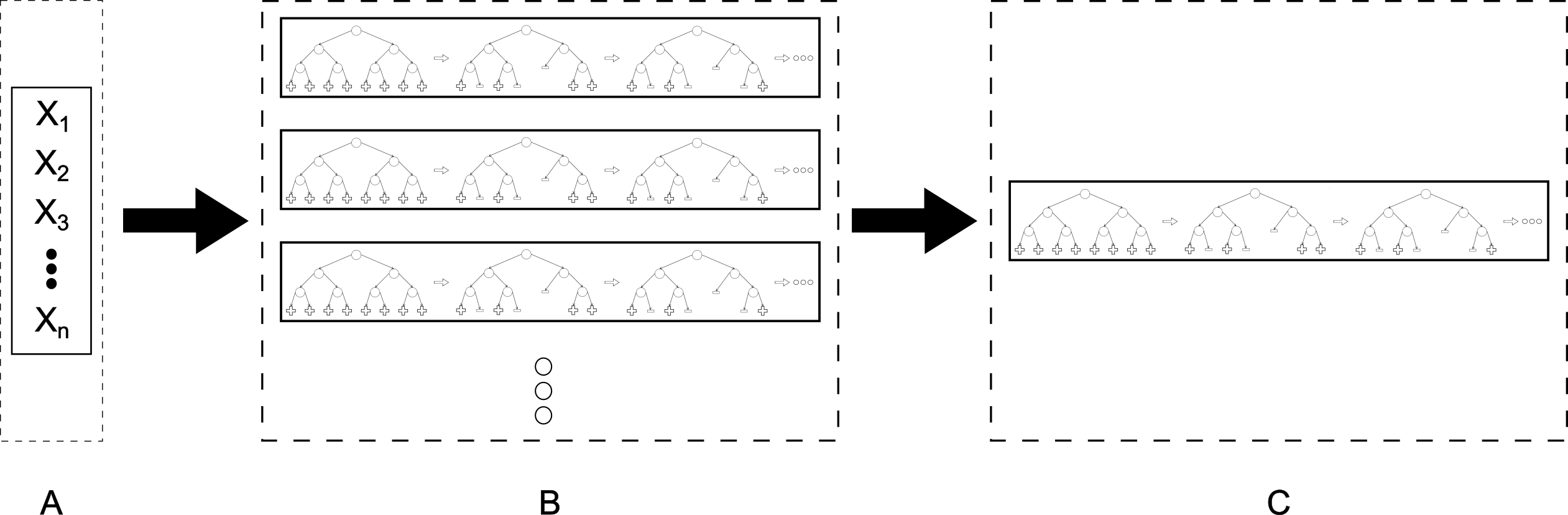}
    \caption{Overall process to obtain the model representative. (A) The parameter set that will be used for training (B) 50 different instances of XGBoost tree were trained to mitigate the randomness influence. Each instance using the AUCPR as metric (C) The instance with higher AUC is chosen as representative to the model.}
    \label{Aprendizado}
\end{figure*}

In order to obtain the optimal performance for the quality analysis\cite{HyperTuning}, each XGBoost model passed a cross-validation process using multiple learning factors and the optimal values were selected using the minimal difference between the mean and the standard deviation of the AUCPR test. Through this method, both the optimal learning factor and the number of rounds associated to them were found to prevent overfitting.

To avoid the randomness influence on the models performance, 50 instances of XGBoost tree were created for each model. All instances were trained with the optimal learning factor, number of rounds, and using the AUCPR as the evaluation metric, as detecting positive results is a higher priority than negative outcomes for a triage application\cite{aucprRef}.

From all the 50 instances in one model, the one with the higher AUC was selected to be the final XGBoost tree for the model. With this selection method, it is possible to have a model that prioritizes the positive outcomes, but without ignoring the prediction of negative outcomes.

\subsection*{Model's performance}

Traditionally, the way to choose the best model is using the area under the receiver operator characteristic (AUROC) as a criterion. However, in view of the fact it is easier for the clinician to interpret a binary output, we used the same principle as in the study published by Pollard \textit{et al}. \cite{ESC} which aimed for maximum sensitivity threshold from the AUROC since the goal was to screen for all the positive cases. Since that study had a bigger population this didn't necessarily implied on a high false positive rate. In our study, due to relatively small number of outcomes in order to have a practical value of sensitivity the choice of 90\% instead of 100\% was necessary so the threshold couldn't be low enough to consider all predictions as positive, overloading the healthcare system. The threshold can be chosen as a bare minimum or as a maximum to the prediction be considered as positive, depending of which case will lead to a higher AUC.
The F2 score was used to compare the models since it is a suitable and effective metric for unbalanced data, especially when the goal is to maximize recall, such as in medical triage applications.
We also checked the importance gain of each parameter used in each model. The Gain implies the relative contribution of the corresponding feature to the model calculated by taking each feature’s contribution for each tree in the model. A higher value of this metric when compared to another feature implies that it is more important to generate a prediction.

\section{Results}

Table \ref{TableOne} presents the demographic characteristics, risk factors, and ECG parameters of the study cohort, stratified by the outcome groups (negative and positive).
The study included 274 patients, with 52.2\% being female. The median BMI was 26.9 kg/m² (IQR: 24.2, 30.4). No significant differences in sex distribution or BMI were observed between the negative and positive outcome groups (p = 0.121 and p = 0.290, respectively). The median age of the cohort was 59.0 years (IQR: 51.0, 67.0) with p=0.001.
A significant difference was observed in the prevalence of previous myocardial MI and previous PCI between the outcome groups. The prevalence of previous MI was significantly higher in the positive outcome group (51.0\%) compared to the negative outcome group (22.9\%) with a p-value of <0.001. Similarly, previous PCI was more common in the positive group (39.2\%) compared to the negative group (9.0\%) with a p-value of <0.001.
\include{Tabelas/TableOne}
Previous cardiac surgery and diabetes also had a higher prevalence in the positive outcome group with 23.5\% versus 8.5\% for the Previous CS and 49\% versus 22.9\% for diabetes with p-values of 0.005 and <0.001 respectively.
No significant differences were found for previous stroke or hypertension between the two groups.
The ECG parameters, including P-wave interval, PR segment interval, QRS interval, QT interval, corrected QTi, and RR interval, showed no significant differences between the negative and positive outcome groups. For instance, the QRS interval was 92.0 ms (IQR: 83.0, 100.6) in the total cohort, with no significant difference between groups (p=0.283).
Among the Global Electrical Heterogeneity (GEH) parameters, significant differences were observed in several measures. The peak QRS-T angle was significantly higher in the positive outcome group (69.9° [IQR: 35.4, 120.4]) compared to the negative group (37.9° [IQR: 25.6, 81.1]) with a p-value of 0.001. The area QRS-T angle also showed a significant difference, being higher in the positive group (92.2° [IQR: 52.7, 137.5]) compared to the negative group (60.4° [IQR: 38.8, 103.3]) with a p-value of 0.007. The peak SVG elevation was significantly greater in the positive group compared to the negative group, with p-values of 0.046 . Additionally, the SVG were significantly different between the groups 65.0 [IQR: 42.4, 85.9] vs 45.6 [IQR: 29.5, 61.9] for the positive group , with p-values of 0.001. This SVG value for the population without events is consistent with the ones published in \cite{haq2021reproducibility} for normal beats of the participants in the Multi-Ethnic Study of Atherosclerosis. 

\begin{table}[H]
\large
\centering
\resizebox{\columnwidth}{!}{%
\begin{tabular}{|c|c|c|c|c|}
\hline
\rowcolor[HTML]{C0C0C0} 
\textbf{Model} & \textbf{F2 Score} & \textbf{AUC (\%)} & \textbf{Sensitivity (\%)} & \textbf{Specificity (\%)} \\ \hline
S   & 0.59 & 59.5 & 94.12 & 20.00 \\ \hline
G   & 0.54 & 54.0 & 94.12 & 3.08 \\ \hline
R   & 0.55 & 62.5 & 94.12 & 6.15 \\ \hline
SRG & 0.62 & 67.6 & 94.12 & 30.77\\ \hline
\end{tabular}
}
\caption{Scores for each one of the built models. Each model used one or more set of the following parameters: Standard ECG parameters (S), GEH calculated (G) and Risk Factors (R). }
\label{Results90}
\end{table}

\begin{figure*}[h!]
    \begin{subfigure}[t]{0.49\textwidth}
        \centering
        \includegraphics[trim=9cm 0cm 8cm 10cm,width=0.3\textwidth,height=0.3\textheight,keepaspectratio]{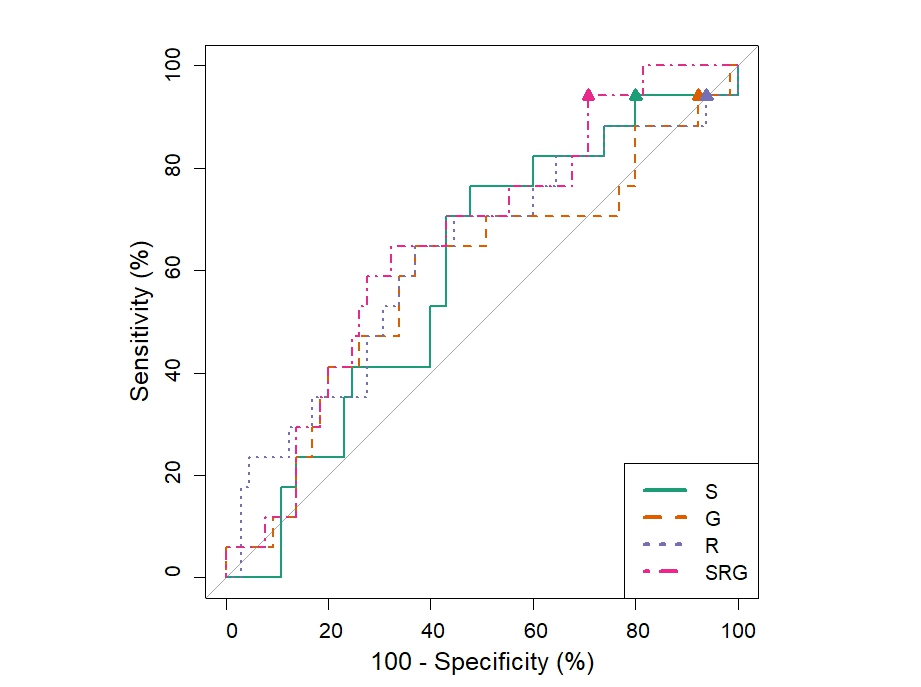}
        \caption{Final AUC of each model as described in \ref{Results90}. S: Standard ECG model, R: Risk factors model, G: Global heterogeneity model, SRG: Combined model with all the S, R and G model features}
        \label{CompROC}
    \end{subfigure}
    \hfill
    \begin{subfigure}[t]{0.49\textwidth}
        \centering
        \includegraphics[trim=4 4 4 4,width=0.95\textwidth,height=1\textheight,keepaspectratio]{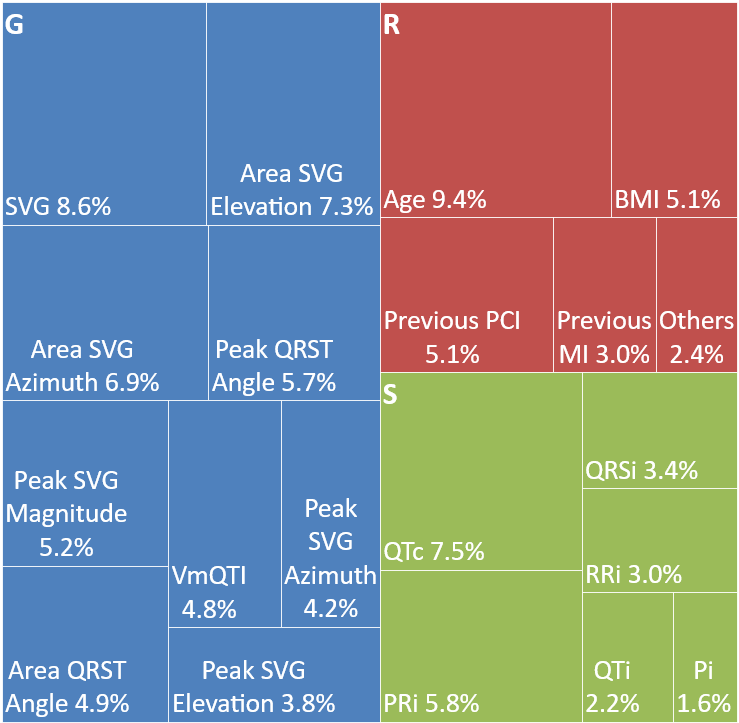}
        \caption{Importance (gain) for all parameters used to build the SRG model}
        \label{ImportanciasSRG}
    \end{subfigure}
    \label{Aprendizado}
    \caption{Comparison of the ROC of the models and importance of the features for the winner model }
\end{figure*}

In Figure \ref{CompROC} and Table \ref{Results90} it is shown the ROCs and the performance metrics of four different machine learning models used for triage, including F2 Score, AUC, sensitivity, and specificity.
 Model SRG achieved the highest F2 score of 0.62, indicating a better balance between precision and recall compared to the other models. Model SRG also had the highest AUC at 67.6\%, suggesting superior overall performance in distinguishing between positive and negative cases. All models maintained the same sensitivity of 94.12\%, indicating a high true positive rate across the board. Model SRG again outperformed the other models with a specificity of 30.77\%, which is significantly higher than Model S (20.00\%), Model R (6.15\%), and Model G (3.08\%).
Overall, Model SRG showed the best performance for triage, with the highest F2 score and AUC, and the best balance between sensitivity and specificity among the models evaluated.
Figure \ref{ImportanciasSRG} show that the results indicate that the winning model (Model SRG) relies heavily on Global Heterogeneity features, with the highest gain importance being attributed to SVG (8.6\%), Area SVG Elevation (7.3\%), and Area SVG Azimuth (6.9\%).Demographics and Risk features, particularly Age (9.4\%), Previous PCI(5.1\%) also have a significant impact on the model. Standard ECG features contribute less to the overall model's performance even though some intervals like with QTc (7.5\%) and PRi (5.8\%) were the third and sixth most important values.

\section{Discussion}
Cardiovascular disease is a significant and costly public health challenge. Determining the appropriate level of healthcare attention for each patient within a high-demand system is a complex task. In this context, the development of computational methods to support medical decisions has shown considerable promise. Besides, given that the referred population already has a history or suspicion of cardiovascular disease, with most ECGs showing abnormalities, it is challenging for clinicians to identify those who genuinely need tertiary care. Therefore, incorporating additional parameters derived from the ECG can significantly improve this triage process in a cost-effective manner.
Implementing these methods using real-world hospital data introduces significant challenges due to data imbalances. Hospital datasets often exhibit a disproportionate representation of certain demographic groups or types of cardiovascular conditions, leading to biased models that may not generalize well to the broader patient population.

The study identified significant differences in several risk factors and GEH parameters between patients with negative and positive outcomes, highlighting the potential importance of these measures in predicting clinical outcomes.
Notably, the present study found that VCG features obtained by post-processing standard ECG signals exhibit higher statistical significance than traditional ECG parameters for triage positive cardiovascular outcomes in a referred population. 

In particular, global heterogeneity parameters demonstrated higher statistical significance compared to standard ECG parameters in this population. Machine learning models using individual parameter types (S, R, and G) revealed that R and G models outperformed the S model. When all parameters were combined, the model achieved a higher Area Under the Curve (AUC) of 76.3\%, an F1 score of 0.48, and a specificity of 49.25\%, while maintaining a fixed sensitivity of 94.12\% across all models.
The choice of a decision tree model fulfilled its purpose of identifying the most contributory features and relating them to clinical practice. The decision points and thresholds used in the model can be clearly visualized, making this machine learning model transparent and allowing for medical scrutiny. This transparency aligns well with findings from other studies, facilitating good clinical agreement.
Notably, the three most important features of the model were related to the SVG. The most critical features included the peak of the SVG magnitude, the SVG itself, and the SVG azimuth calculated using the area. Reinforcing the model results, SVG also showed strong statistical significance (p<0.001), with values of 66.0 [43.4, 86.2] for the population without events and 45.6 [31.4, 64.4] for those who experienced cardiovascular events. 
However, the model's performance can be further enhanced by increasing the sample size and testing its feasibility in clinical settings. Even with the restrictions mentioned, when we use a comparative model it is possible to show that these measurements can improve the triage process pointing the patients that can be referred back with less false positives. This is important to reduce public healthcare costs. 

\section{Conclusion}
In conclusion, incorporating VCG features derived from standard ECG signals significantly enhances the triage and prediction of cardiovascular outcomes in a referred population. Future research should focus on increasing the sample size and conducting feasibility tests in clinical settings to further validate and refine the model. This approach promises to enhance cardiovascular disease management cost-effectively, ultimately improving public health outcomes.

\section{Conflicts of interest and sources of funding}
All authors have disclosed that they do not have any conflicts of interest.
This study was financed by the Funda\c{c}\~{a}o de Amparo \`{a} Pesquisa do Estado de S\~{a}o
Paulo (FAPESP), p  rocess 2021/07005-4

\bibliography{Artigo}

 \end{document}

%% file: Tabelas/TableOne.tex
\begin{table*}[h!]
\centering
\begin{tabular}{|ccccc|}
\hline
\rowcolor[HTML]{999999} 
\multicolumn{1}{|l|}{\cellcolor[HTML]{999999}} &
  \multicolumn{4}{c|}{\cellcolor[HTML]{999999}\textbf{Outcome}} \\ \cline{2-5} 
\rowcolor[HTML]{B7B7B7} 
\multicolumn{1}{|l|}{\multirow{-2}{*}{\cellcolor[HTML]{999999}}} &
  \multicolumn{1}{c|}{\cellcolor[HTML]{B7B7B7}\textbf{Total}} &
  \multicolumn{1}{c|}{\cellcolor[HTML]{B7B7B7}\textbf{Negative}} &
  \multicolumn{1}{c|}{\cellcolor[HTML]{B7B7B7}\textbf{Positive}} &
  \textbf{p-value} \\ \hline
\multicolumn{1}{|c|}{N} &
  \multicolumn{1}{c|}{274} &
  \multicolumn{1}{c|}{223} &
  \multicolumn{1}{c|}{51} &
  \multicolumn{1}{l|}{} \\ \hline
\multicolumn{1}{|c|}{Sex = F} &
  \multicolumn{1}{c|}{143 (52.2)} &
  \multicolumn{1}{c|}{122 (54.7)} &
  \multicolumn{1}{c|}{21 (41.2)} &
  0.112 \\ \hline
\multicolumn{1}{|c|}{Age, y} &
  \multicolumn{1}{c|}{59.0 {[}51.0, 67.0{]}} &
  \multicolumn{1}{c|}{58.0 {[}48.5, 66.0{]}} &
  \multicolumn{1}{c|}{63.0 {[}57.5, 69.5{]}} &
  0.001 \\ \hline
\multicolumn{1}{|c|}{BMI, kg/m²} &
  \multicolumn{1}{c|}{26.9 {[}24.2, 30.4{]}} &
  \multicolumn{1}{c|}{26.5 {[}24.1, 30.5{]}} & 
  \multicolumn{1}{c|}{27.5 {[}25.0, 30.4{]}} & 
   \multicolumn{1}{c|}{0.290} \\ \hline
\rowcolor[HTML]{CCCCCC} 
\multicolumn{5}{|l|}{\cellcolor[HTML]{CCCCCC}\textbf{Risk factors}} \\ \hline
\multicolumn{1}{|c|}{Previous cardiac surgery} &
  \multicolumn{1}{c|}{31 (11.3)} &
  \multicolumn{1}{c|}{19 (8.5)} &
  \multicolumn{1}{c|}{12 (23.5)} &
  0.005 \\ \hline
\multicolumn{1}{|c|}{Previous MI} &
  \multicolumn{1}{c|}{77 (28.1)} &
  \multicolumn{1}{c|}{51 (22.9)} &
  \multicolumn{1}{c|}{26 (51.0)} &
  \textless{}0.001 \\ \hline
\multicolumn{1}{|c|}{Previous PCI} &
  \multicolumn{1}{c|}{40 (14.6)} &
  \multicolumn{1}{c|}{20 (9.0)} &
  \multicolumn{1}{c|}{20 (39.2)} &
  \textless{}0.001 \\ \hline
\multicolumn{1}{|c|}{Previous stroke} &
  \multicolumn{1}{c|}{18 (6.6)} &
  \multicolumn{1}{c|}{13 (5.8)} &
  \multicolumn{1}{c|}{5 (9.8)} &
  0.471 \\ \hline
\multicolumn{1}{|c|}{Hypertension} &
  \multicolumn{1}{c|}{189 (69.0)} &
  \multicolumn{1}{c|}{151 (67.7)} &
  \multicolumn{1}{c|}{38 (74.5)} &
  0.436 \\ \hline
\multicolumn{1}{|c|}{Diabetes} &
  \multicolumn{1}{c|}{76 (27.7)} &
  \multicolumn{1}{c|}{51 (22.9)} &
  \multicolumn{1}{c|}{25 (49.0)} &
  <0.001 \\ \hline
\rowcolor[HTML]{CCCCCC} 
\multicolumn{5}{|l|}{\cellcolor[HTML]{CCCCCC}\textbf{Standard ECG parameters}} \\ \hline
\multicolumn{1}{|c|}{P-wave interval, ms} &
  \multicolumn{1}{c|}{102.2 {[}96.0, 112.0{]}} &
  \multicolumn{1}{c|}{102.0 {[}95.3, 111.2{]}} &
  \multicolumn{1}{c|}{104.0 {[}97.0, 113.7{]}} &
  0.364 \\ \hline
\multicolumn{1}{|c|}{PR segment interval, ms} &
  \multicolumn{1}{c|}{163.0 {[}146.8, 181.7{]}} &
  \multicolumn{1}{c|}{163.0 {[}146.9, 182.3{]}} &
  \multicolumn{1}{c|}{165.0 {[}145.7, 180.7{]}} &
  0.757 \\ \hline
\multicolumn{1}{|c|}{QRS interval, ms} &
  \multicolumn{1}{c|}{92.0 {[}83.0, 100.6{]}} &
  \multicolumn{1}{c|}{91.3 {[}82.7, 100.2{]}} &
  \multicolumn{1}{c|}{92.3 {[}83.8, 104.7{]}} &
  0.283 \\ \hline
\multicolumn{1}{|c|}{QT interval, ms} &
  \multicolumn{1}{c|}{386.0 {[}364.6, 410.1{]}} &
  \multicolumn{1}{c|}{385.3 {[}364.8, 407.5{]}} &
  \multicolumn{1}{c|}{387.3 {[}361.8, 421.0{]}} &
  0.741 \\ \hline
\multicolumn{1}{|c|}{Corrected QTi, ms} &
  \multicolumn{1}{c|}{408.7 (33.7)} &
  \multicolumn{1}{c|}{407.7 (34.1)} &
  \multicolumn{1}{c|}{413.0 (31.8)} &
  0.312 \\ \hline
\multicolumn{1}{|c|}{RR interval, ms} &
  \multicolumn{1}{c|}{920.8 {[}793.7, 1047.9{]}} &
  \multicolumn{1}{c|}{915.3 {[}805.2, 1035.4{]}} &
  \multicolumn{1}{c|}{933.3 {[}745.8, 1075.0{]}} &
  0.744 \\ \hline
\rowcolor[HTML]{CCCCCC} 
\multicolumn{5}{|l|}{\cellcolor[HTML]{CCCCCC}\textbf{GEH Parameters}} \\ \hline
\multicolumn{1}{|c|}{Peak QRST angle,°} &
  \multicolumn{1}{c|}{44.8 {[}25.7, 90.2{]}} &
  \multicolumn{1}{c|}{37.9 {[}25.6, 81.1{]}} &
  \multicolumn{1}{c|}{69.9 {[}35.4, 120.4{]}} &
  0.001 \\ \hline
  \multicolumn{1}{|c|}{Area QRST angle,°} &
  \multicolumn{1}{c|}{66.1 {[}41.2, 111.0{]}} &
  \multicolumn{1}{c|}{60.4 {[}38.8, 103.3{]}} &
  \multicolumn{1}{c|}{92.2 {[}52.7, 137.5{]}} &
  0.007 \\ \hline
\multicolumn{1}{|c|}{Peak SVG Azimuth,°} &
  \multicolumn{1}{c|}{4.4 {[}-6.4, 21.3{]}} &
  \multicolumn{1}{c|}{4.7 {[}-6.5, 19.6{]}} &
  \multicolumn{1}{c|}{4.4 {[}-4.2, 34.1{]}} &
  0.347 \\ \hline
  \multicolumn{1}{|c|}{Area SVG Azimuth,°} &
  \multicolumn{1}{c|}{-11.6 {[}-22.8, 3.8{]}} &
  \multicolumn{1}{c|}{-9.7 {[}-20.8, 4.3{]}} &
  \multicolumn{1}{c|}{-16.1 {[}-24.7, -2.1{]}} &
  0.179 \\ \hline
\multicolumn{1}{|c|}{Peak SVG Elevation,°} &
  \multicolumn{1}{c|}{69.7 {[}60.9, 77.5{]}} &
  \multicolumn{1}{c|}{68.6 {[}60.2, 76.8{]}} &
  \multicolumn{1}{c|}{71.9 {[}64.0, 81.3{]}} &
   0.046 \\ \hline
  \multicolumn{1}{|c|}{Area SVG Elevation,°} &
  \multicolumn{1}{c|}{63.9 {[}54.0, 77.8{]}} &
  \multicolumn{1}{c|}{63.5 {[}55.0, 77.7{]}} &
  \multicolumn{1}{c|}{67.6 {[}47.8, 78.7{]}} &
  0.864 \\ \hline
\multicolumn{1}{|c|}{Peak SVG, mV} &
  \multicolumn{1}{c|}{1.7 {[}1.3, 2.1{]}} &
  \multicolumn{1}{c|}{1.7 {[}1.3, 2.1{]}} &
  \multicolumn{1}{c|}{1.5 {[}1.3, 1.8{]}} &
  {}0.107 \\ \hline
  \multicolumn{1}{|c|}{VmQTI, mVms} &
  \multicolumn{1}{c|}{98.3 {[}78.9, 118.2{]}} &
  \multicolumn{1}{c|}{100.1 {[}77.6, 118.8{]}} &
  \multicolumn{1}{c|}{89.6 {[}80.9, 111.7{]}} &
  {}0.551 \\ \hline
\multicolumn{1}{|c|}{SVG, mV*ms} &
  \multicolumn{1}{c|}{61.1 {[}40.3, 83.7{]}} &
  \multicolumn{1}{c|}{65.0 {[}42.4, 85.9{]}} &
  \multicolumn{1}{c|}{45.6 {[}29.5, 61.9{]}} &
  0.001 \\ \hline  
\end{tabular}%
\caption{Values are n (\%) for categorical variables, median [IQR] for continuos non normal variables and mean (SD) for continuos normal variables. BMI indicates Body Mass Index; MI ,Myocardial Infarction; PCI, Percutaneous Coronary Intervention; SVG, Spatial Ventricular Gradient; VmQTI, Vector Magnitude QT Integral}
\label{TableOne}
\end{table*}